\documentclass[twocolumn, onecolappendix]{aastex701}
\usepackage{amsmath}
\usepackage{enumitem}
\usepackage{newtxtext,newtxmath}
\usepackage{bm}

\hypersetup{breaklinks,colorlinks,citecolor=blue,urlcolor=blue,linkcolor=red}

\newcommand{\uatentry}[2]{\href{http://astrothesaurus.org/uat/#2}{#1 (#2)}}
\newcommand{\Msun}{\,{\rm M}_\odot}

\defcitealias{holm-hansen_catalog_2025}{H25}

\graphicspath{{./}{figures/}}

\shorttitle{\it \textit{StarStream}: Automatic detection algorithm for stellar streams}
\shortauthors{\it Chen et al.}

\begin{document}

\title{\vspace{-6mm}\large \textit{StarStream}: Automatic detection algorithm for stellar streams}

\author[0000-0002-5970-2563]{Yingtian Chen}
\affiliation{Department of Astronomy, University of Michigan, Ann Arbor, MI 48109, USA}
\email{ybchen@umich.edu}

\author[0000-0001-9852-9954]{Oleg Y. Gnedin}
\affiliation{Department of Astronomy, University of Michigan, Ann Arbor, MI 48109, USA}
\email{ognedin@umich.edu}

\author[0000-0003-0872-7098]{Adrian M. Price-Whelan}
\affiliation{Center for Computational Astrophysics, Flatiron Institute, New York, NY 10010, USA}
\email{aprice-whelan@flatironinstitute.org}

\author[0009-0002-1128-2341]{Colin Holm-Hansen}
\affiliation{Department of Astronomy, University of Michigan, Ann Arbor, MI 48109, USA}
\email{cphh@umich.edu}

\correspondingauthor{Yingtian Chen}
\email{ybchen@umich.edu}

\begin{abstract}\noindent 
The \textit{Gaia} mission has led to the discovery of over 100 stellar streams in the Milky Way, most of which likely originated from globular clusters (GCs). As the upcoming wide-field surveys can potentially continue to increase the number of known streams, there is a growing need to shift focus from manual detection of individual streams to automated detection methods that prioritize both quality and quantity. Traditional techniques rely heavily on the visual expectation that GC streams are dynamically cold and thin. This assumption does not hold for all streams, whose morphologies and kinematics can vary significantly with the progenitor's mass and orbit. As a result, these methods are biased toward a subset of the whole stream population, with often unquantified purity and completeness. In this work, we present \textit{StarStream}, an automatic stream detection algorithm based on a physics-inspired model rather than visual expectation. Our method provides a more accurate prediction of stream stars in the multi-dimensional space of observables, while using fewer free parameters to account for the diversity of streams. Applied to a mock GC stream catalog tailored for the \textit{Gaia} DR3 dataset, our algorithm achieves both purity and completeness of at least $65\%$ at Galactic latitudes $|b|>30^\circ$.
\end{abstract}


\section{Introduction}

Stellar streams are elongated tidal structures originating from either an existing or fully dissolved progenitor system, such as a globular cluster (GC) or a dwarf galaxy \citep{lynden-bell_ghostly_1995}. Compared to the high background density of Milky Way (MW) field stars, streams have an extremely low signal-to-noise ratio ($\rm S/N$). As a result, only a few streams were identified prior to the past decade, including the Sagittarius stream by \citet{ibata_dwarf_1994} and the Palomar~5 (Pal~5) stream by \citet{odenkirchen_detection_2001}. These pioneering efforts attempted to increase $\rm S/N$ by selecting a subset of stars that are more likely to belong to streams than to the field, and then visually searching for stream-like structures. This selection was typically performed by applying a matched filter in the color--magnitude diagram, usually a window function centered around the progenitor's isochrone \citep[e.g.,][]{rockosi_matched-filter_2002,grillmair_four_2009,bernard_serendipitous_2014,shipp_stellar_2018}.

The launch of the \textit{Gaia} mission \citep{gaia_collaboration_gaia_2016} has revolutionized the discovery of stellar streams by providing an all-sky map of stars in the six-dimensional phase space, particularly offering high-precision proper motions down to $G\approx20$ since Data Release 2 \citep[DR2,][]{gaia_collaboration_gaia_2018}. This enables additional matched filters based on astrometric measurements, significantly increasing the number of detected streams \citep[see the review by][]{bonaca_stellar_2025}. The all-sky coverage and homogeneity of the \textit{Gaia} data also motivate the development of automatic stream detection methods to replace visual inspection \citep[e.g.,][]{mateu_detection_2011,shih_via_2021}. One such method is \texttt{STREAMFINDER} \citep{malhan_streamfinder_2018}, which uses a mixture model in the multi-dimensional space of observables to automatically detect clusters of stellar orbits within a Gaussian tube. \texttt{STREAMFINDER} successfully identified 87 thin streams in \textit{Gaia} Data Release 3 \citep[DR3,][]{gaia_collaboration_gaia_2023}, including 28 new discoveries \citep{ibata_charting_2024}.

As the number of stream detections grows, more evidence shows that streams have density structure such as \textit{fans} \citep{sesar_evidence_2016}, \textit{gaps} \citep{erkal_number_2016}, \textit{spurs} \citep{price-whelan_off_2018}, and \textit{cocoons} \citep{malhan_butterfly_2019,valluri_gd-1_2024}. These features are likely produced by perturbations in the host galaxy's potential, including bar rotation \citep{hattori_shepherding_2016,price-whelan_spending_2016,pearson_gaps_2017}, disk rotation \citep{nibauer_slant_2024}, and close encounters with other objects \citep{carlberg_pal_2012,ngan_using_2014,erkal_number_2016,erkal_sharper_2017,banik_probing_2018}. On the other hand, recent works proposed that stream density can trace the mass loss history of their progenitors \citep{gieles_supra-massive_2021,chen_stellar_2025}. These breakthroughs emphasize the need to quantify the purity and completeness of stream detection, in order to accurately characterize their density structures.

Previous studies have revealed density structures in individual streams using flexible density models \citep[e.g.,][]{erkal_sharper_2017,tavangar_inferring_2025}. However, these models involve many free parameters, making them computationally expensive when applied to all-sky data and better suited for precisely characterizing stream membership after discovery. Even the simpler model used in \texttt{STREAMFINDER} requires millions of CPU hours on \textit{Gaia} Early Data Release 3 \citep{gaia_collaboration_gaia_2021}. This will greatly limit the usage of these methods for next-generation wide-field photometric surveys, such as those conducted by the \textit{Vera C. Rubin Observatory} \citep[\textit{Rubin},][]{lsst_science_collaboration_lsst_2009} and the \textit{Nancy Grace Roman Space Telescope} \citep[\textit{Roman},][]{spergel_wide-field_2015}.

Furthermore, these models usually represent a stream as a tube surrounding a well-defined track, based on the visual expectation that streams are dynamically cold and thin. However, this assumption is inaccurate, as even GC streams can be dynamically hot or spatially complex depending on the progenitor's mass and orbit \citep{amorisco_feathers_2015}. As a result, such visually inspired models are inefficient at detecting more ``irregular'' streams.

Compared to the visually-inspired models mentioned above, recent theoretical advances in stream formation have enabled physics-inspired models that achieve higher accuracy with fewer free parameters. Specifically, given the host Galactic potential and the progenitor's mass, position, and velocity, particle spray methods \citep{varghese_stellar_2011,lane_tidal_2012,kupper_more_2012,bonaca_milky_2014,fardal_generation_2015,roberts_stellar_2025,chen_improved_2025} can efficiently generate tracer particles that follow the expected distribution of stream stars in six-dimensional phase space, typically requiring only one or even zero free parameters. In particular, the method of \citet{chen_improved_2025} is calibrated to match N-body simulations within 10\% error for typical GC streams, without introducing any additional parameters. As a result, stream models based on this approach can potentially reduce computational cost by a significant amount, while also being able to detect hotter and wider streams.

In this work, we present \textit{StarStream}, an automatic detection algorithm for stellar streams using a physics-inspired stream model based on \citet{chen_improved_2025}. We employ kernel density estimation (KDE) to construct smooth probability density functions (PDFs) for both the stream and background populations. The algorithm is applied to the mock dataset from \citet[][hereafter \citetalias{holm-hansen_catalog_2025}]{holm-hansen_catalog_2025}, tailored for \textit{Gaia} DR3, to quantify its purity and completeness in detecting streams originating from existing GCs. The paper is structured as follows. In \S\ref{sec:method}, we describe the methodology of \textit{StarStream} in detail. We then present validation tests on the mock dataset in \S\ref{sec:validation}. In \S\ref{sec:discussion}, we discuss the motivation for applying the method to upcoming surveys (\S\ref{sec:instruments}) and improvements over other algorithms (\S\ref{sec:improvements}). Finally, we summarize our findings in \S\ref{sec:summary}.

\section{Method}
\label{sec:method}

We distinguish stream members from the background stars using a mixture model, which is a powerful tool for identifying faint structures such as ultra-faint dwarf galaxies \citep[e.g.,][]{pace_proper_2019} and stellar streams \citep[e.g.,][]{malhan_streamfinder_2018,tavangar_inferring_2025}. Specifically, we construct the joint probability density function (PDF) of the stream and background populations as
\begin{equation*}
    p({\bm x}) = f_{\rm s}p_{\rm s}({\bm x}) + (1-f_{\rm s})p_{\rm bg}({\bm x})
\end{equation*}
where $\bm x$ denotes a point in the multi-dimensional observable space, including positions, velocities, colors, magnitudes, and other properties. Traditional methods define PDFs of the stream $p_{\rm s}({\bm x})$ and the background $p_{\rm bg}({\bm x})$ as parametric functions with several fixed or adjustable parameters. The best-fit values of these parameters, together with the stream fraction $f_{\rm s}$, are often estimated by maximizing the log-likelihood,
\begin{equation}
    \ln{\cal L} \equiv \sum_{i=1}^N \ln\left[f_{\rm s} p_{{\rm s},i} + (1-f_{\rm s}) p_{{\rm bg},i}\right]
    \label{eq:lnL}
\end{equation}
where $p_{{\rm s},i} \equiv p_{\rm s}({\bm x}_i)$ and $p_{{\rm bg},i} \equiv p_{\rm bg}({\bm x}_i)$ are the probability densities of the $i$-th star being a member of the stream and background, respectively.

Since parameter estimation is exponentially more computationally expensive as the number of adjustable parameters grows, many methods tend to simplify the stream model by assuming it to be a thin tube along a predefined track. Similarly, the background model is often approximated as uniform or only slowly varying across observables. However, recent advances in the theory of GC stream formation now allow for more accurate stream modeling with even fewer parameters \citep{chen_improved_2025}. Additionally, we can employ a nonparametric KDE to model the nonuniform background without introducing extra model parameters. In this work, we develop a new stream detection method by incorporating these improvements into the mixture model. In the following sections, we detail our approach for accurately estimating $p_{\rm s}({\bm x})$ and $p_{\rm bg}({\bm x})$.

\subsection{Stream probability density}
\label{sec:stream}

To approximate the PDF of streams, we first generate simulated streams around progenitor GCs using the particle spray algorithm. Specifically, we use the \texttt{agama} \citep{vasiliev_agama_2019} implementation of the \citet{chen_improved_2025} algorithm,\footnote{Tutorials for this algorithm are available at \url{https://github.com/ybillchen/particle_spray} and are preserved on Zenodo at \citet{chen_ybillchenparticle_spray_2024}.} which initializes the positions and velocities of stream tracer particles from a multivariate Gaussian distribution, calibrated using N-body simulations of disrupting GCs. This algorithm accurately reproduces the width and length of simulated streams across a wide range of cluster masses and orbital types.

To obtain the probability density in the color--magnitude space, we first assign a stellar mass to each tracer particle by drawing from the \citet{kroupa_variation_2001} initial mass function (IMF). Although the mass function (MF) may evolve due to energy equipartition that preferentially ejects low-mass stars, the high-mass end above \textit{Gaia}'s detection limit ($\gtrsim0.5\Msun$) remains largely consistent with the IMF \citepalias[see][]{holm-hansen_catalog_2025}. We then compute the colors and magnitudes using the MESA Isochrones and Stellar Tracks \citep[MIST,][]{dotter_mesa_2016,mestric_exploring_2022}, taking the progenitor GC's age and metallicity as input. For this study, we adopt \textit{Gaia}'s $G$ magnitude and $\rm BP-RP$ color. In \S\ref{sec:isochrone}, we also test an alternative isochrone model, \texttt{PARSEC} \citep{bressan_span_2012}, which has negligible effect on the final detection quality.

For each stream, we release tracer particles over the last 1~Gyr assuming a uniform ejection rate. The ejection rate can be set time-varying if needed. However, the uniform rate suffices to produce a realistic stream density distribution that is distinguishable from the background, and variations in the ejection rate only slightly affect the density along most streams \citep[see \S2.2 in][]{chen_stellar_2025}. We generate $4000$ tracer particles per stream, which is sufficient to fully sample the multi-dimensional parameter space. The minimum stellar mass used in sampling the mass function is set to the lowest possible mass of the closest tracer particle that remains above the detection limit. Since the heliocentric distances of stream stars vary along the stream, this minimum mass is a conservative choice to ensure that all regions of the stream are well-sampled above the local detection limit.

We use Gaussian KDE to estimate the stream PDF from tracer particles in the $M$-dimensional parameter space, including positions, velocities, colors, magnitudes, and other properties. By denoting the $M$-dimensional coordinate of the $j$-th tracer as ${{\bm x}_j \equiv (x_j^1, x_j^2, \cdots, x_j^M)}$, the probability density at any arbitrary point $\bm x$ is
\begin{align*}
     p_{\rm s}({\bm x}) &\approx \frac{1}{N_{\rm tr}}\sum_{j=1}^{N_{\rm tr}}p_{\rm KDE}({\bm x}|{\bm x}_j,{\bm \sigma}) \\ 
     &\equiv \frac{1}{N_{\rm tr}}\sum_{j=1}^{N_{\rm tr}}\prod_{k=1}^M \frac{1}{\sqrt{2\pi}\sigma_k}\exp\left[-\frac{(x^k-x_j^k)^2}{2\sigma_k^2}\right]
\end{align*}
where $N_{\rm tr}$ is the total number of tracers. It is straightforward to verify that integrating $p_{\rm s}({\bm x})$ over the full $M$-dimensional space yields unity. We define ${\bm \sigma} \equiv (\sigma_1, \sigma_2, \cdots, \sigma_M)$ as the array of KDE kernel bandwidths, with no correlation between each dimension. In practice, we set $\sigma_k$ to 0.1 times the standard deviation of all tracer particles in the $k$-th dimension when $k$ refers to positions or velocities. For magnitudes, we use $\sigma = 0.1$; and for colors, $\sigma = 0.02$. We have verified that varying these values by a factor of $0.5$–$2$ has a negligible effect on our results.

Note that the KDE approach naturally captures the fact that most stream stars are faint, since we sample stream particle masses from the \citet{kroupa_variation_2001} IMF. As a result, more tracer particles are gathered toward the faint end, leading to higher probability densities in that region.

Before applying KDE to estimate the probability density, it is helpful to rotate the equatorial coordinate system so that the new latitude of the stream center is zero. This is particularly important for streams at high declination, where the metric tensor deviates significantly from identity. In this work, we always work in a rotated coordinate frame $(\phi_1, \phi_2)$, where the progenitor GC is located at $(0,0)$ and the proper motion is in the positive $\phi_1$ direction. In this case, the diagonal elements of the metric tensor ${\bf g}={\rm diag}(1,\cos^2\phi_2)$ only deviate from those of the identity tensor by 3\% even with $\phi_2=10^\circ$. This coordinate system is similar to the great circle frame commonly used to describe stellar streams, but is less ambiguous when the stream is wide or has strong curvature.

In practice, the observables of each star often have significant observational uncertainties, denoted by ${\bm \sigma}_0 \equiv (\sigma_{1,0}, \sigma_{2,0}, \cdots, \sigma_{M,0})$, which may exceed the corresponding KDE kernel widths. As we show in Appendix~\ref{sec:kde}, the effective PDF for such a star is the convolution between the original PDF and a Gaussian kernel with standard deviations ${\bm \sigma}_0$. This convolution results in a modified KDE evaluated at the same location, with kernel width replaced by $\sigma_k'^2 \equiv \sigma_k^2 + \sigma_{k,0}^2$. Therefore, for a star with $M$-dimensional coordinates ${\bm x}$ and uncertainties ${\bm \sigma}_0$, the stream probability density becomes
\begin{equation*}
    p_{\rm s}({\bm x},{\bm\sigma}_0) \approx \frac{1}{N_{\rm tr}}\sum_{j=1}^{N_{\rm tr}}p_{\rm KDE}({\bm x}|{\bm x}_j,{\bm \sigma}')
\end{equation*}
where ${\bm \sigma}'$ incorporates both the KDE kernel width and observational uncertainty.

For \textit{Gaia} data specifically, the uncertainties in positions and magnitudes are almost always smaller than the corresponding KDE bandwidths. For simplicity and computational efficiency, we therefore ignore the uncertainties in these parameters in our subsequent analysis and consider only the uncertainties in proper motions and color.

It is worth noting that \textit{Gaia} astrometric uncertainties have nonzero correlations. These correlations influence the error propagation from the original equatorial frame to the rotated $(\phi_1, \phi_2)$ frame. In Appendix~\ref{sec:astrometry}, we explicitly calculate the linear uncertainty propagation associated with this coordinate transformation. However, to simplify our calculations, we do not include these correlations when performing the convolution with the Gaussian kernel, allowing us to treat each dimension independently during KDE evaluation. This simplification has only a minor effect on the inferred probability density, as the correlations are generally weak ($r < 0.5$).

\subsection{Background probability density}
\label{sec:background}

Similarly to the stream probability density, we also use Gaussian KDE to estimate the background PDF around a stream. We directly use the observed stars in the same spatial region as the stream to construct the KDE estimator. However, there are typically $10^7$ observed stars in these regions. Constructing and evaluating a multi-dimensional Gaussian KDE with so many data points can be extremely inefficient. To address this, we perform an initial selection around the isochrone at the progenitor's distance to exclude stars that are too red or too blue to be plausible stream members. For \textit{Gaia} specifically, we select stars within a color offset $\Delta({\rm BP-RP}) = 0.5$ around the main sequence, red-giant branch (RGB), and horizontal branch of the isochrone. We also extend the isochrone with $\Delta G = 1.5$ above the tip of the RGB and around the horizontal branch to include stars clustered in those regions. This selection is rather conservative, given that the typical spread around the isochrone is $\sigma_{\rm BP-RP} \lesssim 0.1$ \citep{riello_gaia_2021} even considering the distance spread of stream stars. Nevertheless, it still reduces the number of background stars by a factor of $2-10$.

To further speed up the calculation, we use a grid interpolation technique, given the fact that the background population is relatively homogeneous and uncorrelated across most observables. For \textit{Gaia}, we account for correlations only in the two-dimensional position space ${\bm x}_{\rm pos}^{\rm 2D}$ and the two-dimensional color–magnitude space ${\bm x}_{\rm cm}^{\rm 2D}$. The proper motions $\mu_{\phi1} \equiv \dot{\phi}_1 \cos\phi_2$\footnote{Some works refer to $\dot{\phi}_1 \cos\phi_2$ as $\mu_{\phi1}^*$. We use $\mu_{\phi1}$ for simplicity, as this does not lead to confusion.} and $\mu_{\phi2} \equiv \dot{\phi}_2$ are assumed to be uncorrelated with other observables. Under these assumptions, the background PDF becomes
\begin{equation*}
    p_{\rm bg}({\bm x}) = p_{\rm bg}^{\rm pos}({\bm x}_{\rm pos}^{\rm 2D}) \, p_{\rm bg}^{\rm \mu_{\phi1}}(\mu_{\phi1}) \, p_{\rm bg}^{\rm \mu_{\phi2}}(\mu_{\phi2}) \, p_{\rm bg}^{\rm cm}({\bm x}_{\rm cm}^{\rm 2D})
\end{equation*}
where ${\bm x} \equiv \left({\bm x}_{\rm pos}^{\rm 2D}, \mu_{\phi1}, \mu_{\phi2}, {\bm x}_{\rm cm}^{\rm 2D}\right)$. We evaluate the PDF in each subspace independently using Gaussian KDE on a rectangular grid large enough to cover all background stars in that subspace. Each subspace is kept at most two-dimensional, since the number of grid points grows exponentially with dimensionality. For efficiency, we randomly select $10^4$ background stars near the stream to construct the KDE estimators. We adopt bandwidths of $0.5^\circ$ for position, $1\ {\rm mas\,yr^{-1}}$ for proper motions, and $0.1$ for the color–magnitude space. The grid spacings are set equal to the corresponding bandwidths. The final results are not sensitive to the exact choice of bandwidths or grid spacing, as these values sufficiently resolve the density structure in observable space. Quantitatively, we have verified that multiplying or dividing the bandwidths and grid spacings by a factor of 2 results in only $\lesssim10\%$ variation in detection purity and completeness for a typical stream.

Note that the PDF estimation can be biased in regions with $\phi_2 > 10^\circ$, where the metric tensor deviates from the identity tensor by more than 3\% (see \S\ref{sec:stream}). Our stream generation method may also deviate from the actual stream track in regions far from the GC if the adopted Galactic potential model is inaccurate. For these reasons, the KDE approach is best suited for relatively small regions, such as a $10^\circ$ cone around the GC. Nevertheless, this region is still sufficiently large to enclose the half-number radius for $2/3$ of the simulated streams.

Finally, we perform linear interpolation on the grids and compute the product of the independent PDFs to estimate the background probability density $p_{\rm bg}$ at any point of interest. Unlike our estimation of the stream probability density, where the simulated stream has no observational uncertainty, the real \textit{Gaia} data already include measurement uncertainties. As a result, the background PDFs obtained above are already the convolution of the original PDFs with Gaussian kernels characterizing uncertainties. Therefore, we do not need to apply the convolution again when evaluating the background PDF at a given point. This simplification is particularly helpful, as performing convolution would be highly inefficient within the grid interpolation framework.

An alternate method is first deconvolving the real \textit{Gaia} data to obtain the actual underlying PDF of the background, and then using the same approach as our stream model to compute the PDF. The first step can be achieved using techniques such as ``extreme deconvolution'' \citep{bovy_extreme_2011}. Although this method provides a coherent approach to obtain PDFs for both the stream and the background, it is more computationally expensive by $\sim1000$ times compared to grid interpolation as this alternate method requires performing the full KDE evaluation over background stars.

Although we demonstrate our method using \textit{Gaia} data as an example, our method can be readily adapted to other datasets. For instance, metallicity and radial velocity can be included as additional observables when dealing with spectroscopic surveys. Since we perform grid fitting separately for most dimensions, adding extra observables only increases the computational cost linearly.

\subsection{Stream detection}

We optimize our mixture model for the stream and background populations by varying the stream fraction $f_{\rm s}$ to maximize the log-likelihood function in Eq.~(\ref{eq:lnL}). The best-fit stream and background probability densities for star $i$ are then given by $f_{\rm s}p_{{\rm s},i}$ and $(1 - f_{\rm s})p_{{\rm bg},i}$, respectively. Following the standard definition used in mixture models, the membership probability that star $i$ belongs to the stream is given by
\begin{equation}
    P_{{\rm s},i}\equiv\frac{f_{\rm s}p_{{\rm s},i}}{f_{\rm s}p_{{\rm s},i}+(1-f_{\rm s})p_{{\rm bg},i}}.
\end{equation}
We consider stars with membership probability greater than a chosen threshold $P_{\rm th}$ to be identified as stream members. In this work, we adopt the standard setup of the mixture model with $P_{\rm th} = 0.5$. However, we emphasize that $P_{\rm th}$ is an adjustable parameter that can be tuned depending on whether the analysis prioritizes completeness or purity.

\section{Method validation}
\label{sec:validation}

We validate our new method by applying it to a mock catalog of stream and background stars tailored for \textit{Gaia} DR3, based on the \citetalias{holm-hansen_catalog_2025} stream catalog.

\subsection{Mock observational data}
\label{sec:mock_data}

Our mock catalog consists of both a stream population and a background population, each described by six observables: positions $(\phi_1, \phi_2)$, proper motions $(\mu_{\phi1}, \mu_{\phi2})$, color ($\rm BP-RP$), and magnitude ($G$). We do not include radial velocity or metallicity, as these quantities are available for only a small subset of stars in \textit{Gaia} DR3. Parallax is also excluded because of its large uncertainty at the typical heliocentric distances of stellar streams.

The stream population is from the mock catalog of GC streams in a simulated MW-like galaxy (ID 523889) by \citetalias{holm-hansen_catalog_2025}. This catalog generates synthetic streams around a mock GC population based on the GC formation model of \citet{chen_catalogue_2024}. It fits the host potential of simulated galaxies at each snapshot using basis function expansion (BFE), accounting for the time evolution of the potential by linearly interpolating between snapshots. The catalog then explicitly integrates the orbit of each GC in this potential over the last $\sim$3.5~Gyr and computes the mass loss rate based on the GC mass and the local tidal field. \citetalias{holm-hansen_catalog_2025} initializes each GC with the \citet{kroupa_variation_2001} IMF and releases stars according to the time-varying mass loss rate. Stars are released probabilistically, with the ejection probability inversely proportional to the square root of stellar mass. The released stars form stellar streams using the particle spray algorithm by \citet{chen_improved_2025}. The catalog provides the initial mass, age, and metallicity $\rm [Fe/H]$ of each star, allowing us to assign synthetic \textit{Gaia} photometry directly using the MIST isochrone model.

Ideally, we should generate our simulated streams in the same potential used in \citetalias{holm-hansen_catalog_2025} for full consistency. However, because it is challenging to constrain the time evolution of the Galactic potential in practice, we use only the static potential at the final snapshot to avoid over-idealization. 

The stream duration in \citetalias{holm-hansen_catalog_2025} is longer than that of most observed GC stream segments \citep[$\lesssim 1$~Gyr,][]{chen_stellar_2025}. In this work, we only use the portions of streams that were released in the last 1~Gyr to mimic streams that are currently observable. Since the main goal of this method is to detect streams around GCs, it is also unnecessary to include stream stars released more than 1~Gyr ago or located outside the $10^\circ$ cone centered on the GC, as these stars tend to be more sensitive to the time evolution of the Galactic potential.

To create a more realistic stream population, we also add observational errors to the mock stream stars. Similarly to \S\ref{sec:stream}, we neglect errors in positions and magnitudes. For proper motion uncertainties, we adopt the following parametric form from \textit{Gaia}'s performance website,\footnote{\url{https://www.cosmos.esa.int/web/gaia/science-performance}} which describes the dependence of the uncertainty on apparent magnitude,
\begin{equation}
    \sigma_\mu = \frac{\sqrt{40 + 800z + 30z^2}}{1000}\ {\rm mas\,yr^{-1}}
    \label{eq:sigma_mu}
\end{equation}
where $z=10^{0.4[\max(G,13)-15]}$. This expression increases from $0.01$ to $0.6\ {\rm mas\,yr^{-1}}$ with $G=13-20$. These errors can be significant, especially given that the intrinsic proper motion spread of a Pal~5-like stream is much smaller.

\textit{Gaia}'s performance website suggests multiplying Eq.~(\ref{eq:sigma_mu}) by a fudge factor of $1.03$ and $0.89$ for the proper motions of right ascension and declination, respectively. In the rotated frame $(\phi_1,\phi_2)$, we have verified that simply setting this factor to unity for both coordinates also reproduce the actual proper motion uncertainties with sufficient accuracy. We then add Gaussian noise to the original proper motions using Eq.~(\ref{eq:sigma_mu}).

Similarly, we parameterize the dependence of color uncertainty on apparent magnitude using the following expression,
\begin{equation}
    \sigma_{\rm BP-RP} = 10^{[\max(G,14)-23]/3}
    \label{eq:sigma_color}
\end{equation}
which increases from $0.001$ to $0.1$ for $G=14-20$. This expression accurately reproduces the mean $\rm BP-RP$ color uncertainty in \textit{Gaia} DR3 by \citet{riello_gaia_2021}.

For the background population, we directly use observational data from \textit{Gaia} DR3 within the same $10^\circ$ cone centered on the progenitor GC of each stream in the \citetalias{holm-hansen_catalog_2025} catalog. We only select stars with $G < 20$. Based on the \textit{Gaia} DR3 selection function from \texttt{gaiaunlimited} \citep{cantat-gaudin_empirical_2023}, we find that the mean completeness exceeds $99\%$ in regions where streams are located. Even for the most incomplete case, the completeness remains above $90\%$. Thus, this magnitude cut provides near-complete coverage for our mock dataset. We also apply the same color--magnitude selection as in \S\ref{sec:background} to reduce the number of background stars that are extremely unlikely to be misidentified by the method.

This results in between 1 million (near the Galactic pole) and 30 million (near the Galactic center) background stars in a selected region, compared to $10-10,000$ stream stars in the same region. This dramatic contrast between the two populations (up to $\sim4$ orders of magnitude) underscores the challenge of detecting streams using traditional methods.

We restrict our analysis to streams originating from surviving GCs with $M>10^3\Msun$ and containing at least 10 stars after applying the above selection criteria. This leads to 158 streams from our chosen catalog from \citetalias{holm-hansen_catalog_2025}. For each stream, we construct a mock dataset following the procedure described above and evaluate our detection algorithm on it, assuming no prior knowledge of how the dataset is constructed.

\subsection{Method performance}
\label{sec:performance}

\begin{figure*}
    \centering
    \includegraphics[width=\linewidth]{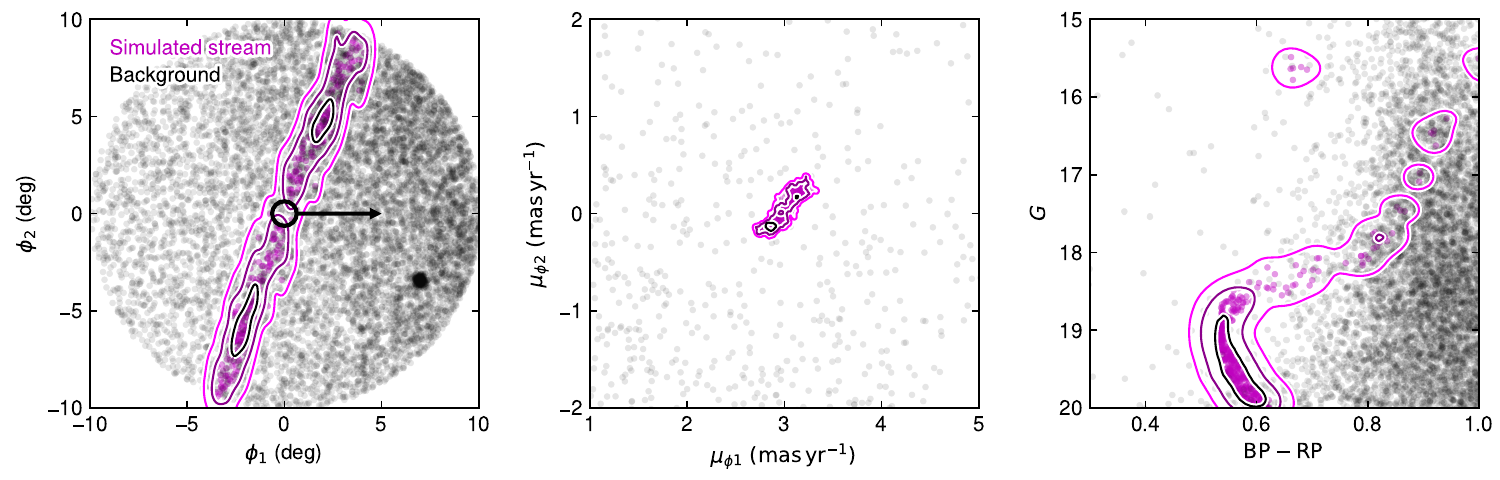}
    \includegraphics[width=\linewidth]{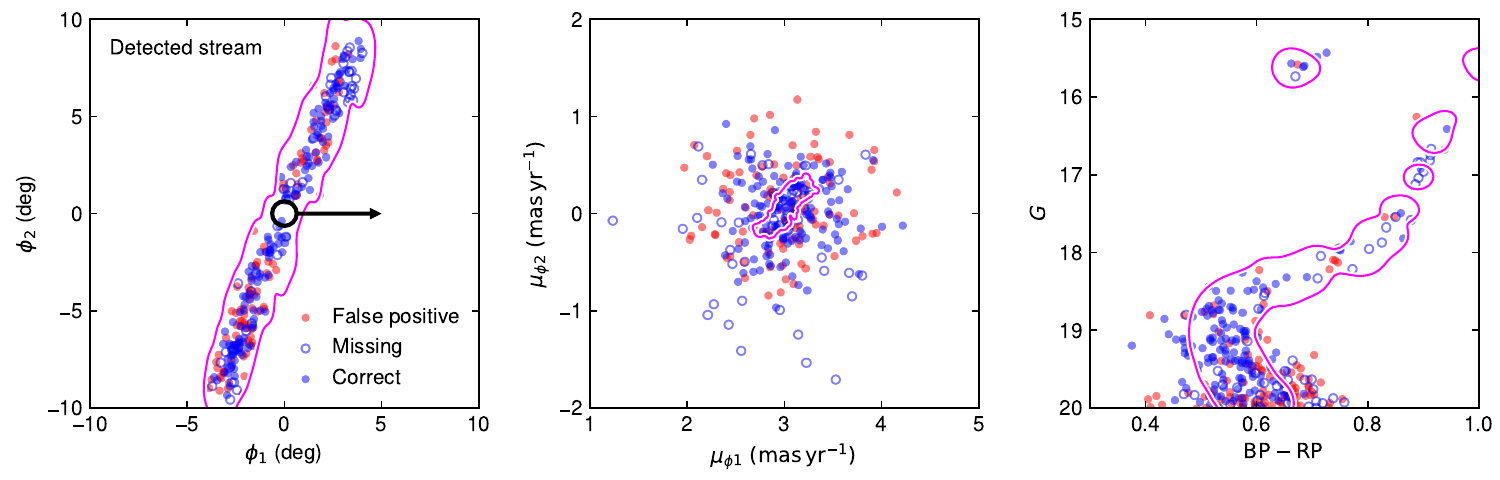}
    \vspace{-5mm}
    \caption{Demonstration of a test of the method on a mock stream. \textit{Top row:} distributions of simulated tracer particles (magenta) and background stars (black) in position space (\textit{left}), proper motion space (\textit{middle}), and color--magnitude space (\textit{right}). We plot the stream PDF $p_{\rm s}({\bm x})$ from Gaussian KDE as gray contours. The three contours from dark to light represent three values of $\ln p_{\rm s}({\bm x})-\ln p_{\rm s,max}({\bm x})=-0.5, -2, -4.5$, corresponding to the 1-$\sigma$, 2-$\sigma$, and 3-$\sigma$ ranges of the standard Gaussian distribution, respectively. We only show $10^4$ background stars randomly chosen from the total of $4\times 10^6$ for visual clarity. \textit{Bottom row:} application of the method on this mock stream using $P_{\rm th}=0.4$. We show stars that are false positives (red circle), missed by the method (blue open circle), or correctly detected (blue solid circle) in the same subspaces as the \textit{top row}. The mock stream members already have uncertainties added and are mixed with background stars from real \textit{Gaia} DR3 data. We also plot the contours of $\ln p_{\rm s}({\bm x})-\ln p_{\rm s,max}({\bm x})=-4.5$ for comparison. For the \textit{left column} of both rows, we show the location of the progenitor GC as the circle of tidal radius. The velocity of the GC is represented by the arrow.}
    \vspace{2mm}
    \label{fig:method_demo}
\end{figure*}

First, we illustrate key concepts of our method by applying it to an example mock stream originating from a GC with mass $M=5.6\times10^5\Msun$, located at a Galactocentric radius $r_{\rm Gal} = 11$~kpc and a Galactic latitude $b = 21^\circ$. The tidal radius is $125$~pc, corresponding to an angular size of $0.63^\circ$ at its heliocentric distance $d_\odot=8.7$~kpc. In the \textit{top row} of Fig.\ref{fig:method_demo}, we show the distributions of simulated tracer particles in position space, proper motion space, and color–magnitude space. The stream PDF $p_{\rm s}({\bm x})$ is constructed via KDE using these simulated tracers. Although we display $p_{\rm s}({\bm x})$ as 2D contours in each of the three subspaces, we emphasize that the KDE is constructed in the full six-dimensional space of all observables and projected onto 2D subspaces.

We then apply our method to the mock dataset, which includes 194 mock stream stars from \citetalias{holm-hansen_catalog_2025} and approximately four million background stars from \textit{Gaia} DR3. We detect 244 stream members, of which 140 are true members (see the \textit{bottom row} of Fig.\ref{fig:method_demo}). Most of the false positives and missed detections are either along the RGB, where background contamination is high, or near $G = 20$, where observational uncertainties become significant according to Eqs.(\ref{eq:sigma_mu}) and (\ref{eq:sigma_color}). As we show later in \S\ref{sec:instruments} and Fig.~\ref{fig:sn}, both of these regions have lower $\rm S/N$ than the main sequence turnoff. We also note that our simulated stream in Fig.\ref{fig:method_demo} samples the RGB less densely than the main sequence due to the lower abundance of RGB stars. This, however, only weakly affects the detection quality, as RGB stars only contribute a small fraction of the total and are intrinsically hard to detect in any case due to their low $\rm S/N$.

It is worth noting that for faint stars ($G \gtrsim 19$), the uncertainties in color and proper motions can exceed the intrinsic spread of the entire stream. Our method accounts for this by convolving the KDE with a Gaussian blob centered on each star, as described in \S\ref{sec:stream}. Without this convolution, many stars would be shifted outside the effective selection window in proper motion and color space, leading to significantly fewer detections than expected. In contrast, while our detection accuracy does decrease for these faint stars, the method still tends to recover the correct total number of stream members by balancing false positives and missed detections. This is expected as our mixture model tends to reproduce the correct number of stream stars to recover the correct density ratio between the stream and the background.

\begin{figure}
    \centering
    \includegraphics[width=\linewidth]{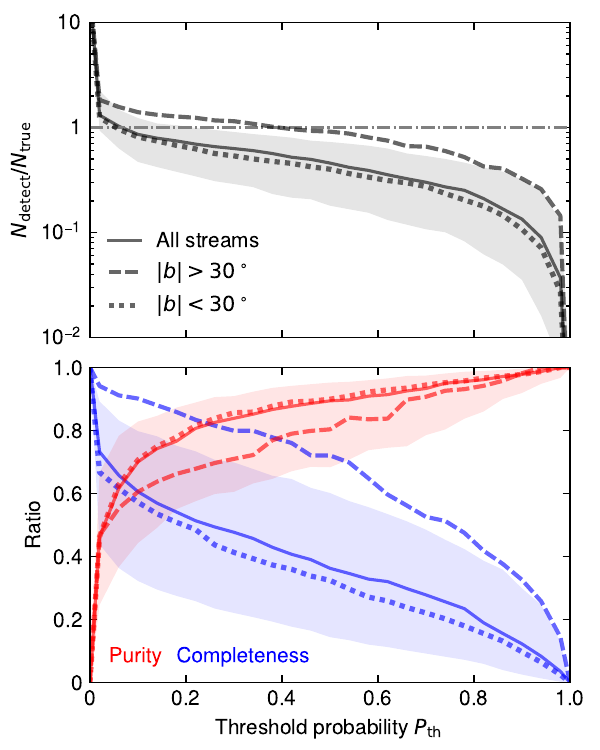}
    \caption{Detection ratio (\textit{upper panel}) and completeness/purity (\textit{lower panel}) of detected stream members as functions of probability threshold. The solid lines stand for the median values among all test streams, while the shaded ranges show the interquartile ranges. Detection ratio $=1$ is highlighted as the dot-dashed line. We also show the median detection ratio, completeness, and purity for streams with progenitor GCs at low Galactic latitude ($|b|<30^\circ$, dotted curves) and high Galactic latitude ($|b|>30^\circ$, dashed curves) separately. Since the purity at $P_{\rm th}=1$ is not well defined, we extrapolate the values at $P_{\rm th}=0.99$ out to 1 for visual clarity.}
    \label{fig:fdetect_completeness_purity_vs_pth}
\end{figure}

Next, we examine the statistical performance of our method by applying it to all mock streams. We quantify detection quality using three metrics: detection ratio, purity, and completeness. The \emph{detection ratio} is defined as the ratio of the total number of detected stream members $N_{\rm detect}$ to the total number of true members $N_{\rm true}$. In the \textit{upper panel} of Fig.~\ref{fig:fdetect_completeness_purity_vs_pth}, we show $N_{\rm detect}/N_{\rm true}$ as a function of the threshold probability $P_{\rm th}$. Starting from a large value $\gg 1$ at $P_{\rm th} = 0$, the median detection ratio of all streams rapidly drops to 1 at $P_{\rm th} \approx 0.07$, with an interquartile range of approximately $0.3$~dex. It then gradually declines to 0.2 at $P_{\rm th} = 0.8$, followed by a rapid decrease to 0 as $P_{\rm th}$ approaches 1.

Furthermore, we define the \textit{purity} as the ratio of correctly identified stream members $N_{\rm correct}$ to the total number of detected members $N_{\rm detect}$. The \textit{completeness} is defined as the ratio of $N_{\rm correct}$ to $N_{\rm true}$. By definition, both purity and completeness range from 0 to 100\%, and the ratio between completeness and purity equals the detection ratio. In the \textit{lower panel} of Fig.~\ref{fig:fdetect_completeness_purity_vs_pth}, we show the median purity and completeness as functions of $P_{\rm th}$. The median purity increases rapidly from 0 to about 70\% for $P_{\rm th} = 0-0.1$, and then gradually approaches 100\% as $P_{\rm th} \rightarrow 1$ (note that purity is not well defined at $P_{\rm th} = 1$). In contrast, completeness drops from 100\% to around 60\% over $P_{\rm th} = 0-0.1$, and then continues to decrease to 0 at $P_{\rm th}=1$. The two curves intersect at $P_{\rm th} \approx 0.07$, where the detection ratio also reaches unity.

When divided by Galactic latitude $b$, the high-latitude streams with $|b| > 30^\circ$ reach a detection ratio of unity at $P_{\rm th} =0.5$. However, a broad range of $P_{\rm th} = 0.2-0.6$ yields detection ratios that deviate from unity by less than $0.2$~dex. For these high-latitude streams, purity and completeness also intersect at $P_{\rm th} \approx 0.5$, reaching 80\% and 72\% respectively. On the other hand, the majority of streams are located at low Galactic latitudes $|b| < 30^\circ$. Therefore, their detection quality more closely follows the overall trend of the full sample, which is worse than the high-latitude streams. This difference comes from the distinct background densities between the two populations. Compared to the high-$b$ streams near the Galactic poles, the low-$b$ streams closer to the Galactic plane are contaminated by roughly 10 times more background stars. In these regions, the signal from actual stream stars is more easily washed out by the strong background noise, leading to a lower probability to be identified as stream members.

Based on the above tests, our default $P_{\rm th}=0.5$ yields high purity and completeness $>70\%$ for streams with relatively low contamination at $|b| > 30^\circ$. The similarity between purity and completeness also ensures that $N_{\rm detect}$ serves as an unbiased estimator of $N_{\rm true}$. This is particularly important for inferring properties of the progenitor GC, such as the mass loss rate. Streams at $|b| < 30^\circ$ should be treated more carefully sincem the median completeness can drop to about 40\%. However, we emphasize that $P_{\rm th}$ is a user-defined parameter that can be set anywhere between 0 and 1, depending on whether the application prioritizes purity or completeness.

\begin{figure}
    \centering
    \includegraphics[width=\linewidth]{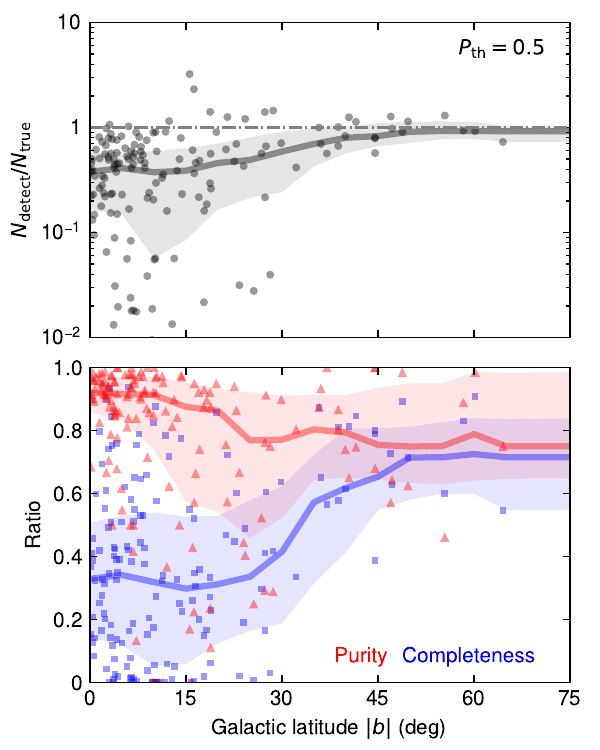}
    \caption{Detection ratio (\textit{upper panel}) and completeness/purity (\textit{lower panel}) of detected stream members as functions of the progenitor GC's absolute Galactic latitude $|b|$ with $P_{\rm th}=0.5$. Individual streams are shown as black circles (detection ratio), blue diamonds (completeness), and red triangles (purity). The solid lines stand for the median values among all test streams, while the shaded ranges show the interquartile ranges. Detection ratio $=1$ is highlighted as the dot-dashed line. We calculate the percentiles at any $b$ using nearby streams smoothed by the Gaussian kernel, with bandwidth varying linearly from $5^\circ$ to $15^\circ$ from the Galactic plane to the poles.}
    \label{fig:fdetect_completeness_purity_vs_b}
\end{figure}

To further study the dependence of detection quality metrics on Galactic latitude, we plot them against $|b|$ in Fig.~\ref{fig:fdetect_completeness_purity_vs_b}. We apply Gaussian kernel smoothing \citep[similarly to \S3.2 in][]{chen_formation_2023} to estimate the median and interquartile ranges of the three metrics as smooth functions of $|b|$, using a kernel bandwidth that grows linearly from $5^\circ$ to $15^\circ$ over the range $|b| = 0-75^\circ$. We find that the detection ratio $N_{\rm detect}/N_{\rm true}$ remains consistent with unity for $|b| > 30^\circ$. However, the ratio deviates unity near the Galactic plane, where the dispersion is also larger since $\sim20\%$ of streams with detection ratios $<0.1$ are all located at $|b|<30^\circ$.

The new method reports zero detections for only 4\% of all streams. This suggests that the method almost always identifies some stream members as long as a stream is present. However, this conclusion requires further validation by addressing two key questions: 1) How often does the method report false detections when no stream is present? 2) To what extent does the idealization of our mock data benefit detection quality? We explore these questions in the following subsections.

\subsection{Null test}
\label{sec:null}

In addition to the above tests that focus on how many stream member stars can be correctly detected, it is also important to perform the null test to ensure that the method does not report false detections when there is no stream. Therefore, we design the null test by removing the signal (i.e., mock stream stars) from the test dataset and applying the method only to the background stars in the same region around each GC. A perfect method should yield zero detections in the null test. However, due to random fluctuations in the background that are not captured by the KDE, it is possible that a small fraction of background stars are accidentally better described by the stream model. This can lead to the detection of ``false streams''.

\begin{figure}
    \centering
    \includegraphics[width=\linewidth]{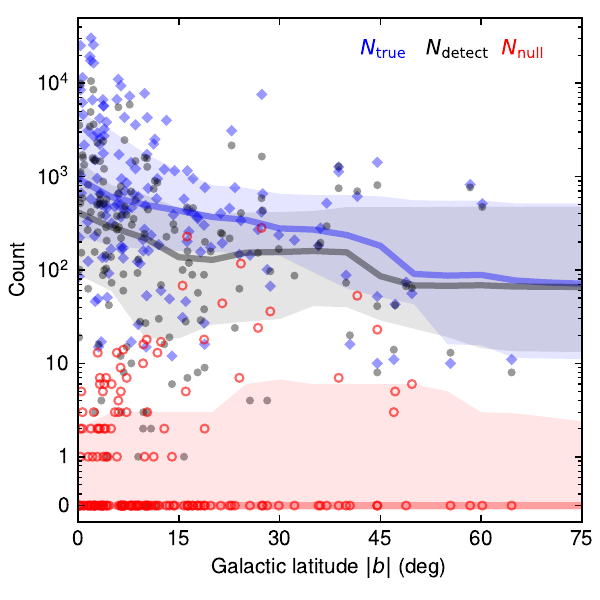}\\
    \vspace{-3mm}
    \caption{Number of detected stars in the null test $N_{\rm null}$ (red open circles) as a function of the progenitor GC's absolute Galactic latitude $|b|$. For comparison, we also show the true number of stream stars $N_{\rm true}$ (blue solid diamonds) and the number of detections $N_{\rm detect}$ (black solid circles) without removing the signal. Similarly to Fig.~\ref{fig:fdetect_completeness_purity_vs_b}, we show the the median values as solid lines, while the shaded ranges represent the interquartile ranges.
    }
    \label{fig:null_test}
\end{figure}

In Fig.~\ref{fig:null_test}, we show the number of detections in the null test, $N_{\rm null}$, as a function of Galactic latitude $|b|$. For comparison, we also plot the true number of stream stars $N_{\rm true}$ and the number of detections $N_{\rm detect}$ when the signal is included. Our method reports exactly zero detections for 60\% of the mock GCs. For the remaining cases, $N_{\rm null}$ is still generally much smaller than the corresponding $N_{\rm true}$ and $N_{\rm detect}$. Only 11\% of cases yield $N_{\rm null} \geq 10$. Conversely, 16\% of cases have $N_{\rm detect} < 10$. Based on this, we recommend excluding detections below a threshold $N_{\rm detect} \approx 10$ to avoid most false positives while not discarding too many true stream members.

\subsection{Dependence on isochrone models}
\label{sec:isochrone}

It should be noted that the \citetalias{holm-hansen_catalog_2025} catalog uses the same MIST isochrone model adopted in this work to generate the the mock photometry. Before applying our method to real observational data, it is important to assess how sensitive the detection quality is to the choice of isochrone model. To quantify this, we perform a test where we switch to the \texttt{PARSEC} isochrones \citep{bressan_span_2012} for constructing the simulated stream PDF. Specifically, we set $\rm [M/H]$ in \texttt{PARSEC} equal to the metallicity $\rm [Fe/H]$ of the mock GC and fix the age to 10~Gyr. These settings are intentionally approximate to mimic real observational conditions, where GC ages are often uncertain.

\begin{figure*}
    \centering
    \includegraphics[width=0.49\linewidth]{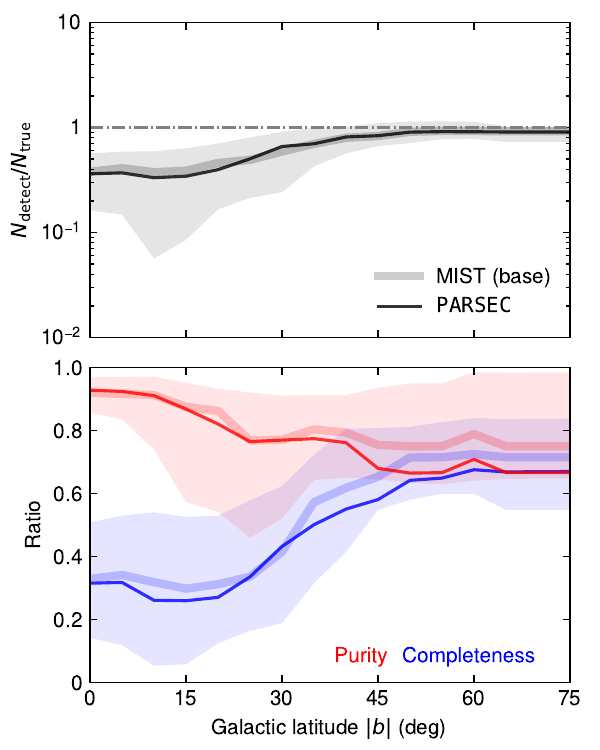}
    \includegraphics[width=0.49\linewidth]{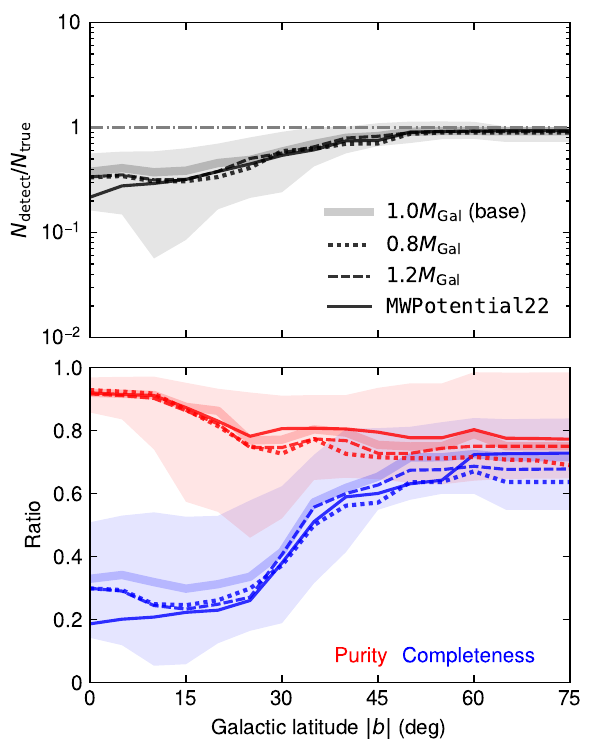}
    \caption{Same as Fig.~\ref{fig:fdetect_completeness_purity_vs_b}, but with the alternative isochrone model $\texttt{PARSEC}$ (\textit{left column}) and three alternative Galactic potential models (\textit{right column}): 1) the base potential model scaled down by $20\%$ (dotted curves), 2) the base potential model scaled up by $20\%$ (dashed curves), and 3) the \texttt{MilkyWayPotential2022} model from \texttt{gala}. We show the interquartile ranges for the base case for comparison.}
    \label{fig:fdetect_completeness_purity_vs_b_dependence}
\end{figure*}

In the \textit{left column} of Fig.~\ref{fig:fdetect_completeness_purity_vs_b_dependence}, we show the detection ratio, purity, and completeness as functions of $|b|$ when switching to the \texttt{PARSEC} isochrones. Despite using a different isochrone model with approximate parameters, these metrics show no statistically significant differences compared to those in Fig.\ref{fig:fdetect_completeness_purity_vs_b}. This is because our KDE-based approach provides sufficient flexibility to accommodate variations between isochrone models. Furthermore, the variance introduced by changing isochrone models is negligible compared to the typical color dispersion from observational errors and magnitude dispersion due to the spread in heliocentric distances. Thus, our method is robust to the choice of isochrone model.

\subsection{Dependence on Galactic potential models}

Our simulated streams are generated in the same Galactic potential model as the final snapshot corresponding to our selected catalog from \citetalias{holm-hansen_catalog_2025}. When applied to real data, however, we do not know the Galactic potential exactly. Most existing Milky Way potential models predict a total mass differing by $\lesssim 10\%$ \citep[see the review by][]{hunt_milky_2025}. To quantify the impact of using an inaccurate potential, we test three alternative models. The first two are based on the same BFE framework as \citetalias{holm-hansen_catalog_2025}, but with all expansion coefficients scaled by $\pm20\%$. This results in a proportional change in the enclosed mass at all radii. We refer to these cases as $0.8M_{\rm Gal}$ and $1.2M_{\rm Gal}$. The third model is \texttt{MilkyWayPotential2022} from \texttt{gala} \citep{price-whelan_gala_2017,price-whelan_adrngala_2024}, which has been validated against MW mass measurements out to $\sim150$~kpc \citep{hunt_milky_2025}.

The \textit{right column} of Fig.~\ref{fig:fdetect_completeness_purity_vs_b_dependence} shows the detection ratio, purity, and completeness as functions of $|b|$ for these models. In almost all cases, these detection metrics decrease only slightly. An exception is \texttt{MilkyWayPotential2022}, which shows a $\lesssim10\%$ increase in purity at $|b| > 60^\circ$. This is likely a stochastic effect given the small number of streams at such high latitudes. All metrics remain within the interquartile range (which is even narrower than the $1\sigma$ range) of the original values, indicating that even a $20\%$ variation in the potential has little effect on detection quality. Since recent measurements of the MW potential have uncertainties of only $\approx 10\%$ \citep[e.g.,][]{ibata_charting_2024}, these tests are conservative.

The \texttt{MilkyWayPotential2022} model performs almost identically to the original BFE model. This is encouraging as the former is designed to match the real MW instead of the simulated galaxy. While both models have similar halo structures beyond 10~kpc, their disk components differ significantly, with enclosed masses near 1~kpc differing by $\approx 30\%$. This suggests that the performance of our method is insensitive to the exact choice of Galactic potential model within $10^\circ$ around the GC.

We emphasize, however, that our method has not been tested beyond $10^\circ$, where the Galactic potential becomes more important. Even within $10^\circ$, the slight decrease in purity for the alternative models is largely driven by stars located between $5^\circ$ and $10^\circ$, indicating that the influence of the potential grows with distance from the progenitor. Moreover, the LMC may introduce a larger perturbation to the MW's potential \citep[see review by][]{vasiliev_effect_2023} than that in the simulated galaxy. Since the \citetalias{holm-hansen_catalog_2025} catalog does not include a galaxy with a realistic LMC analog, we are unable to quantify the effect of LMC in this work.

\subsection{Dependence on stream generation algorithms}

The mock stream catalog also uses the same \citet{chen_improved_2025} particle spray algorithm as adopted in this work. To further validate our method, we conduct an additional test by generating a mock stream using an N-body simulation, which is generally considered more accurate than most particle spray methods. Specifically, we initialize the simulation using the same initial conditions as the example stream in \S\ref{sec:performance} and Fig.~\ref{fig:method_demo}. Following \citet{chen_improved_2025}, we model the progenitor GC using the \citet{king_structure_1966} model with $W = 8$, a typical value for Galactic GCs. We set the particle mass to $10\Msun$, the softening length to $1$~pc, and the simulation time step to $2^{-13}\ {\rm kpc\,km^{-1}\,s}\approx0.1$~Myr. We then backtrack the GC orbit for 1~Gyr in the same static Galactic potential described in \S\ref{sec:mock_data}, and run the N-body simulation forward to the present day using the fast-multipole gravity solver \texttt{falcON} \citep{dehnen_very_2000,dehnen_hierarchical_2002}. Although the N-body simulation includes only collisionless dynamics and omits close stellar encounters, it is sufficiently distinct from the particle spray model to serve our purpose of validating the detection method with an alternative algorithm.

Since the particles in the simulation are different from individual stars, we randomly select a subsample of escaped particles and re-sample their masses using the stellar mass distribution of stars with $G < 20$ from \citetalias{holm-hansen_catalog_2025}. We then assign $G$ magnitudes and $\rm BP-RP$ colors to these particles using the MIST isochrone. This ensures that the new mock stream from the N-body simulation has the same number of stars and mass distribution as its counterpart in \citetalias{holm-hansen_catalog_2025}, which is important for a fair comparison of detection quality metrics.

Finally, we add background stars to the new mock stream following \S\ref{sec:mock_data} and apply our stream detection method. Among the 194 mock stream stars, we detect 230 members, with 133 correct detections. These numbers are only about $5-10\%$ lower than those reported in \S\ref{sec:performance}. This slight decrease is expected, as we attempt to recover stream members generated using a different algorithm. However, since our method evaluates membership probability by comparing the stream probability density to the background, the detection performance is not strongly sensitive to the specific stream generation algorithm as long as the $\rm S/N$, defined as the ratio between the densities of actual stream members and background stars in the multi-dimensional observable space, is greater than 1.

\subsection{Influence of dust extinction}

When applying \textit{StarStream} to real data, it is important to account for dust extinction, particularly near the Galactic plane where the color excess reaches $E(B-V) > 1$. Extinction reduces the number of observable stream stars above the detection limit by up to a factor of 10 near the mid-plane, while also shifting the stream to redder colors, where the background density is higher (see the \textit{upper right} panel of Fig.~\ref{fig:method_demo}). In this section, we investigate the influence of these effects by incorporating dust extinction into our mock dataset.

We use the Python package \texttt{dustmaps} \citep{green_dustmaps_2018} to compute the extinction $A_V$ of each stream star in \citetalias{holm-hansen_catalog_2025}, based on the map of \citet[][hereafter SFD]{schlegel_maps_1998}. This map is recalibrated by \citet{schlafly_measuring_2011} with $R_V = 3.1$. The $A_V$ values are then passed to the MIST bolometric correction interpolation table\footnote{\url{https://waps.cfa.harvard.edu/MIST/model_grids.html}} to obtain the $G$-band extinction and $\rm BP-RP$ color excess. Since the table only covers $A_V = 0-6$, we remove stars with $A_V > 6$, as they are also likely too faint to be observable. Note that the SFD map provides extinction along the full line of sight from the solar system to infinity. This overestimates the true extinction, since streams are located at finite distances. While the overestimation is probably modest, this test should be regarded as an extreme case that assumes maximum possible extinction.

The detection quality would decline significantly if we directly apply the same method to the new dataset, since the simulated isochrone no longer aligns with the reddened distribution of stream stars in the color--magnitude space. Fortunately, in practice we have access to extinction values for most MW GCs from catalogs such as \citet{harris_catalog_1996}, allowing us to account for realistic extinction when simulating the stream. This is equivalent to replacing the original isochrone with a reddened one, where the extinction and color excess are calculated from the progenitor's $A_V$. Here, we take each GC's $A_V$ directly from the SFD map at its location. As before, we exclude streams whose progenitor GCs have $A_V > 6$. Since extinction makes streams fainter, the number of streams with at least 10 stars also decreases, leaving 123 valid streams out of the original 158.

Stars from the same stream may have different $A_V$ values in the mock dataset, since extinction is not constant within the search radius. However, we redden the simulated isochrone using only a single $A_V$ value at the center of the stream. This reflects the practical difficulty of obtaining precise extinction for every individual stream star. We emphasize that our tests are designed to reproduce the actual detection quality of \textit{StarStream} when applied to real data. Thus, it is important to avoid any over-idealization.

\begin{figure}
    \centering
    \includegraphics[width=\linewidth]{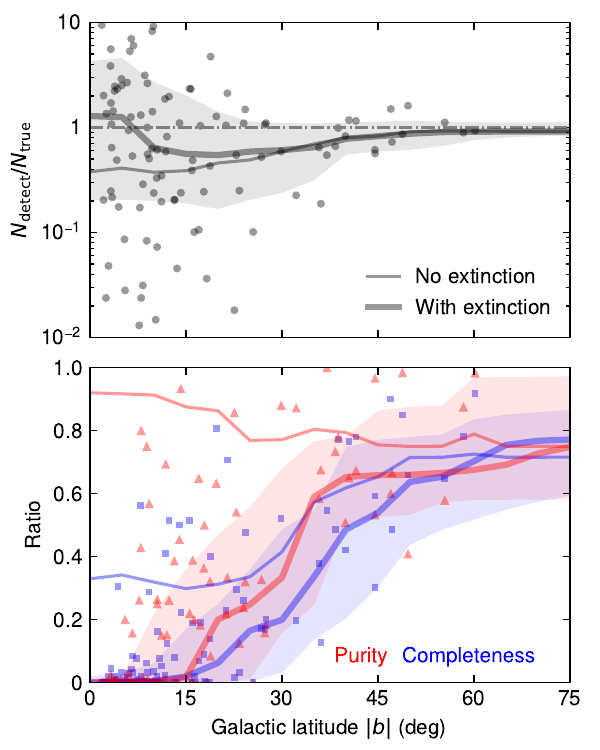}
    \caption{Same as Fig.~\ref{fig:fdetect_completeness_purity_vs_b}, but including dust extinction from the SFD map. We also show the original cases (no extinction) as thin curves for comparison.}
    \label{fig:fdetect_completeness_purity_vs_b_extinction}
\end{figure}

In Fig.~\ref{fig:fdetect_completeness_purity_vs_b_extinction}, we show the detection ratio, completeness, and purity after applying \textit{StarStream} to the new mock dataset that includes extinction. As expected, both completeness and purity decrease substantially near the Galactic plane ($|b| < 30^\circ$). Close to the mid-plane, purity falls from $\sim 90\%$ to below 10\%, and completeness drops to a similar level. This decline is primarily due to the high background density near the reddened isochrone and the lower number of stream members that remain observable. We find that completeness can be improved by lowering the probability threshold to $P_{\rm th} < 0.1$, whereas varying $P_{\rm th}$ between $0.01-0.99$ does not significantly improve purity. In contrast, the high-latitude regions are only slightly affected. For streams at $|b| > 30^\circ$, the median completeness and purity decrease by only $\sim 10\%$ to 62\% and 67\%, respectively. The detection ratio in this latitude range is nearly unchanged.

\begin{figure}
    \centering
    \includegraphics[width=\linewidth]{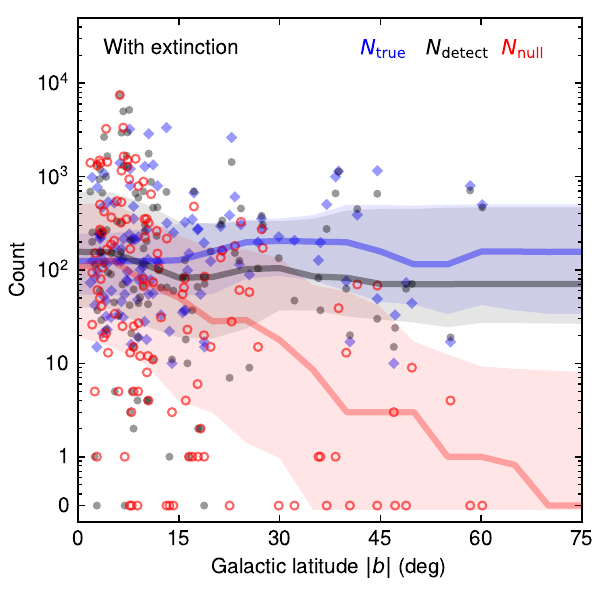}\\
    \vspace{-3mm}
    \caption{Same as Fig.~\ref{fig:null_test}, but including dust extinction from the SFD map.}
    \label{fig:null_test_extinction}
\end{figure}

We also perform the null test described in \S\ref{sec:null} on the new dataset, and show it in Fig.~\ref{fig:null_test_extinction}. In this case, only 13\% of detections yield $N_{\rm null} = 0$. This reduction is mainly due to the higher false detection rate at low latitudes, where $N_{\rm null}$ is nearly the same as $N_{\rm detect}$ for $|b| < 15^\circ$. However, the high-latitude region again remains almost unaffected. At $|b| > 30^\circ$, false and true detections can still be cleanly separated using the same threshold of $N_{\rm detect} \approx 10$ recommended in \S\ref{sec:null}.

Therefore, although both completeness and purity of \textit{StarStream} decrease at low latitudes when accounting for extreme extinction, the method still achieves high values $\approx 65\%$ at $|b| > 30^\circ$, where extinction is less significant. Since the extinction adopted in this section overestimates the true values, these results represent lower limits of the actual detection quality. As discussed later in \S\ref{sec:instruments}, future spectroscopic and deep photometric surveys are necessary to reveal low-$b$ streams with high extinction.

\bigskip
\section{Discussion}
\label{sec:discussion}

\subsection{Application to other surveys}
\label{sec:instruments}

Although this work is framed with Gaia DR3, it can be straightforwardly extended to other surveys. In this subsection, we discuss the importance of spectroscopic surveys and deep photometric surveys in further enhancing detection quality.

Our method operates in the six-dimensional space of positions, proper motions, color, and magnitude. We find that the median detection ratio drops significantly to 18\% when color and magnitude are excluded. In this case, the median purity and completeness also decrease to 50\% and 6\%, respectively. If proper motions are excluded, we cannot even detect any stars for more than 80\% of streams. These results demonstrate that six-dimensional information of positions, proper motions, colors, and magnitudes is important for recovering most streams in \textit{Gaia} DR3. For streams with extremely low density, strong background contamination, or high extinction we may need even more independent observables. This highlights the importance of spectroscopic surveys, such as the Apache Point Observatory Galactic Evolution Experiment \citep[APOGEE,][]{majewski_apache_2017}, the Southern Stellar Stream Spectroscopic Survey \citep[S$^5$,][]{li_southern_2019}, and the Dark Energy Spectroscopic Instrument \citep[DESI,][]{desi_collaboration_overview_2022} Milky Way Survey \citep[MWS,][]{cooper_overview_2023}, which provide radial velocities and metallicities and may help reveal very faint streams.

\begin{figure}
    \centering
    \includegraphics[width=\linewidth]{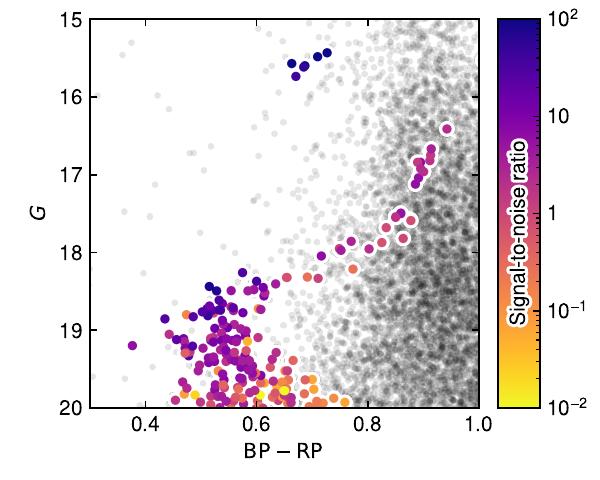}
    \vspace{-6mm}
    \caption{Distribution of $\rm S/N$ in the color--magnitude diagram for actual members of the mock stream in Fig.~\ref{fig:method_demo}. The signal and noise densities are estimated using a similar KDE method in \S\ref{sec:stream} and \S\ref{sec:background}, respectively. We multiply the stream kernel width by a factor of 3 to account for the lower number density of stars that do not sample the parameter space as well as simulated tracers. The same $10^4$ background stars as in Fig.~\ref{fig:method_demo} are also shown for reference.}
    \label{fig:sn}
\end{figure}

Our example stream in Fig.~\ref{fig:method_demo} shows that most correctly identified members lie near the main sequence turnoff, where the $\rm S/N$ is high. In Fig.~\ref{fig:sn}, we present the distribution of $\rm S/N$ in the color–magnitude diagram, where $\rm S/N$ is defined as the ratio between the densities of actual stream members and background stars. We estimate both densities in the six-dimensional space using KDE approaches similar to those described in \S\ref{sec:stream} and \S\ref{sec:background}. However, we multiply the stream kernel widths in \S\ref{sec:stream} by a factor of 3, since the number of member stars is smaller than the number of simulated tracer particles.

It is remarkable that a significant number of stars have $\rm S/N$ values larger than unity in six-dimensional space, despite the number of stream members being orders of magnitude smaller than the background contaminants. The region with $\rm S/N > 1$ coincides with the area of high purity and completeness near the main sequence turnoff at $G \approx 19$. Brighter stars in the RGB suffer from higher background contamination, resulting in lower $\rm S/N$. Stars in the horizontal branch show relatively high $\rm S/N$ because they are bluer than most background stars; however, the horizontal branch contributes only a few members to the stream. The majority of stars lie below the main sequence turnoff and have lower $\rm S/N$ due to large observational uncertainties, as described by Eqs.~(\ref{eq:sigma_mu}) and (\ref{eq:sigma_color}). This issue is not unique to the example stream. Since the typical main sequence turnoff is at absolute magnitude $M_G = 3-4$, corresponding to $G = 20$ at a heliocentric distance $\approx 20$~kpc, \textit{Gaia} can barely detect stars fainter than the turnoff for most MW streams. This highlights the importance of deep photometric surveys such as LSST and \textit{Roman}, which are expected to detect significantly more stream members thanks to their much deeper detection limits \citep{pearson_forecasting_2024,holm-hansen_catalog_2025}. Moreover, \textit{Roman} can also provide high-quality astrometry measurements for faint stars, significantly improving $\rm S/N$ below the main sequence turnoff.

\subsection{Improvements to existing methods}
\label{sec:improvements}

The new detection method constructs stream KDE via a particle spray algorithm, which only requires the progenitor GC's mass, position, and velocity as input. We then assign colors and magnitudes to the tracers based on an isochrone model, which depends only on the progenitor's metallicity and age. In practice, most of these input parameters are available from existing catalogs \citep[e.g.,][]{hilker_galactic_2019}. The only exception is age, which, however, has a weak impact on our results (see \S\ref{sec:isochrone}). Therefore, our method avoids making unnecessary and unrealistic assumptions about the stream's morphology and kinematics. Additionally, we also construct the background KDE directly from a subsample of observed stars.

As a result, both the stream and background models have no free parameters. The mixture model that combines these two components includes only one free parameter: the stream fraction $f_{\rm s}$. This minimal parameterization offers a direct advantage in computational efficiency. On average, our Python implementation takes $\sim10$ minutes to detect a single stream when running on 32 cores of an Intel Haswell CPU. The total computation time to analyze all mock streams is approximately $1000$ CPU hours of computation time, which is orders of magnitude faster than typical mixture models. For instance, \texttt{STREAMFINDER} requires millions of CPU hours on a similar dataset \citep{ibata_charting_2021}.

Although the particle spray method used here takes no free parameters, our stream model is more accurate than simply assuming the stream is an elongated structure along its orbital track. A representative example of the latter is \texttt{STREAMFINDER}, which detects clustering of stellar orbits within a Gaussian tube. Using \texttt{STREAMFINDER}, \citet{ibata_charting_2024} successfully detected 16 streams originating from known GCs in \textit{Gaia} DR3. To compare performance, we apply our method to the same 16 GCs in \textit{Gaia} DR3, using GC properties from the \citet{hilker_galactic_2019} catalog and the Galactic potential model \texttt{MilkyWayPotential2022}. Excluding M68, $\omega$Cen, and M5, whose streams in \citet{ibata_charting_2024} are not connected to the progenitors, all other streams extend at least $5^\circ$ within our 1~Gyr integration time. In this region, our method detects on average 5 times as many stream members. Even for the Pal~5 stream, which is widely thought to be one of the most complete, we detect 131 members compared to 76 reported in \citet{ibata_charting_2024}, which a meaningful improvement. Notably, the actual Pal~5 stream extends beyond $5^\circ$ because it formed over approximately 6~Gyr \citep{chen_stellar_2025}, much longer than our integration time. For the remaining streams that extend farther within our 1~Gyr integration time, we detect on average 4 times as many member stars inside $10^\circ$.

\textit{StarStream} is not the first physics-motivated attempt to search for GC streams. For example, \citet{grillmair_extended_2022,yang_spectacular_2023,grillmair_multiple_2025} have successfully identified tidal features around individual GCs by comparing observations with simulated streams. Compared to these works, our approach automates this technique using KDE. In addition, we provide quantitative metrics for detection quality, demonstrating the broader potential of this method for identifying more GC streams.

\section{Summary}
\label{sec:summary}

In this work, we present \textit{StarStream}, an automatic detection algorithm for stellar streams based on a physics-inspired stream model. We construct a mixture model in the multidimensional space of observables, including positions, velocities, colors, magnitudes, etc. The model consists of background and stream components, whose PDFs are represented using KDE. For the background, we build the KDE from a subsample of observed stars; for the stream, we construct the KDE from tracers generated using the particle spray algorithm of \citet{chen_improved_2025}. We illustrate the method using an example stream in Fig.~\ref{fig:method_demo}.

We quantitatively assess the detection quality of our method around existing GCs using the mock stream catalog from \citetalias{holm-hansen_catalog_2025}, which is tailored to \textit{Gaia} DR3 and includes six observables: sky coordinates ($\phi_1,\phi_2$), proper motions ($\mu_{\phi1},\mu_{\phi2}$), color ($\rm BP-RP$), and magnitude ($G$). Our mock dataset incorporates magnitude-dependent uncertainties for each observable, and we include all \textit{Gaia} DR3 stars as the background population. The method achieves both purity and completeness around $65\%$ even with extreme dust extinction ($>70\%$ without extinction). The detection ratio is near unity for high-latitude streams (Fig.~\ref{fig:fdetect_completeness_purity_vs_b} and Fig.~\ref{fig:fdetect_completeness_purity_vs_b_extinction}). For low-latitude streams, however, high background contamination and extinction can significantly reduce both purity and completeness to $<10\%$.

Next, we perform a series of tests to examine the robustness of the method. We begin with a null test to evaluate the frequency of false positive detections. After removing the signal (i.e., mock stream stars) from the dataset, our method correctly reports $N_{\rm null}=0$ for $13$\% of all streams with extreme extinction ($60\%$ without extinction), while a threshold $N_{\rm detect} \approx 10$ cleanly separates true and false detections for high-latitude streams (Fig.~\ref{fig:null_test} and Fig.~\ref{fig:null_test_extinction}). We further test the method using a different isochrone model and different Galactic potential models  (Fig.~\ref{fig:fdetect_completeness_purity_vs_b_dependence}), and a different stream generation algorithm. These alternative configurations do not significantly weaken the detection quality. Being robust to alternate isochrone models is important in application to real data as the predicted isochrone can vary significantly among different models.

We find that both purity and completeness drop significantly when proper motions or color and magnitude are excluded from the input dataset. This emphasizes the importance of incorporating multiple independent observables for stream detection. With the full six-dimensional input, however, the $\rm S/N$ can exceed unity even when the number of stream members is orders of magnitude smaller than the background contaminants (Fig.~\ref{fig:sn}). Stars with high $\rm S/N$ are primarily located near the main sequence turnoff, coinciding with those that exhibit high purity and completeness. In contrast, fainter stars near $G = 20$ and brighter stars on the RGB both have lower $\rm S/N$ due to large observational uncertainties or strong background contamination, respectively.

Finally, we compare our new method to existing methods such as \texttt{STREAMFINDER}. Our method is several orders of magnitude more computationally efficient, primarily because the physics-inspired stream model requires no free parameters. This greatly accelerates the model optimization process. At the same time, for streams associated with existing GCs in \citet{ibata_charting_2024}, our method detects on average 5 times as many member stars within the $5^\circ$ circle. It is also worth noting that the method may uncover additional streams when applied to the full set of GCs in \textit{Gaia} DR3.

We have published the package \textit{StarStream} on GitHub via \url{https://github.com/ybillchen/StarStream}, where we also provide example Python notebooks for running the code. The code requires the mass and six-dimensional phase-space coordinates of the progenitor to generate stream tracers using the particle spray algorithm. It also requires an isochrone that is fit to the progenitor to compute mock photometry for the tracers. The input dataset should be a multi-dimensional array of observables, coupled with another array of observational uncertainties of the same shape. Users can also specify the threshold probability $P_{\rm th}$, tracer particle ejection rate, KDE kernel widths, interpolation grid spacings for the background PDF, and the Galactic potential, if different values from the default of this paper are preferred.

\section*{Acknowledgments}
We thank Monica Valluri, Eric Bell, Katya Gozman, and Jacob Nibauer for insightful discussions. 
YC, OYG, and CHH were supported in part by National Aeronautics and Space Administration through contract NAS5-26555 for Space Telescope Science Institute programs HST-AR-16614 and JWST-GO-03433.
This research benefited from the Gravity in the Local Group conference hosted by the McWilliam's Center for Cosmology and Astrophysics, Carnegie Mellon University.

\software{
\texttt{agama} \citep{vasiliev_agama_2019}, 
\texttt{numpy} \citep{harris_array_2020}, 
\texttt{matplotlib} \citep{hunter_matplotlib_2007}, 
\texttt{scipy} \citep{virtanen_scipy_2020}, 
\texttt{astropy} \citep{the_astropy_collaboration_astropy_2018}, 
\texttt{gala} \citep{price-whelan_gala_2017,price-whelan_adrngala_2024}, 
\texttt{pandas} \citep{the_pandas_development_team_pandas-devpandas_2024},
\texttt{dustmaps} \citep{green_dustmaps_2018},
\texttt{falcON} \citep{dehnen_very_2000,dehnen_hierarchical_2002},
\texttt{gaiaunlimited} \citep{cantat-gaudin_empirical_2023}
}

\appendix
\vspace{-6mm}

\section{Kernel density estimation with uncertainties}
\label{sec:kde}

Given a sample of $N$ points $\{x_j\}$, we can estimate the probability density function $p(x)$ at any point of interest using Gaussian KDE,
\begin{align*}
    p(x) &\approx \frac{1}{N}\sum_{j=1}^Np_{\rm KDE}(x|x_j,\sigma) \\
    &\equiv\frac{1}{N}\sum_{j=1}^N\frac{1}{\sqrt{2\pi}\sigma}\exp\left[-\frac{(x-x_j)^2}{2\sigma^2}\right].
\end{align*}
However, if the location of the point of interest is uncertain and follows a distribution function $f(x)$, the expected probability density $p$ is the average of $p(x)$ weighted by $f(x)$,
\begin{equation*}
    p = \int_{-\infty}^\infty p(x)f(x)dx \approx \sum_{j=1}^N \int_{-\infty}^\infty p_{\rm KDE}(x|x_j,\sigma)f(x)dx.
\end{equation*}
If we assume $f(x)$ is a Gaussian function centered at $x_0$ with uncertainty $\sigma_0$, the last integral is the convolution of two Gaussian distributions ${\cal N}(x_i,\sigma_i^2)\star {\cal N}(0, \sigma_0^2)$ with the independent variable $x_0$. This convolution is simply another Gaussian distribution ${\cal N}(x_i,\sigma'^2)\equiv{\cal N}(x_i,\sigma^2+\sigma_0^2)$. We can thus obtain the expected probability density as
\begin{equation*}
    p \approx \frac{1}{N}\sum_{j=1}^Np_{\rm KDE}(x|x_j,\sigma').
\end{equation*}
Therefore, the expected probability density of a point with Gaussian uncertainty $\sigma_0$ equals the standard Gaussian KDE at the same location with bandwidth replaced by $\sigma'^2\equiv\sigma^2+\sigma_0^2$ for every sample point $\{x_i\}$.

\section{Astrometry uncertainty propagation}
\label{sec:astrometry}

The uncertainties of astrometry measurements in sky coordinate system $(\alpha,\delta,\mu_\alpha,\mu_\delta,\varpi,v_r)$ are commonly quantified as a $6\times6$ covariance matrix
\begin{equation*}
    {\bf C} = \left(
    \begin{array}{cccc}
        V_\alpha & C_{\alpha\delta} & \cdots & C_{\alpha v_r} \\
        C_{\alpha\delta} & V_\delta & \cdots & C_{\delta v_r} \\
        \vdots & \vdots & \ddots & \vdots \\
        C_{\alpha v_r} & C_{\delta v_r} & \cdots & V_{v_r}
    \end{array}
    \right)
\end{equation*}
where $V_i=\sigma_i^2$ is the variance of quantity $i$ and $C_{ij}=\sigma_i\sigma_j\rho_{ij}$ is the covariance between quantities $i$ and $j$, in which $\rho_{ij}=\rho_{ji}$ is the correlation coefficient.

To obtain the covariance matrix in the rotated frame $(\phi_1,\phi_2,\mu_{\phi1},\mu_{\phi2},\varpi',v_r')$,
\begin{equation*}
    {\bf C}' = \left(
    \begin{array}{cccc}
        V_{\phi_1} & C_{\phi_1\phi_2} & \cdots & C_{\phi_1 v_r'} \\
        C_{\phi_1\phi_2} & V_{\phi_2} & \cdots & C_{\phi_2 v_r'} \\
        \vdots & \vdots & \ddots & \vdots \\
        C_{\phi_1 v_r'} & C_{\phi_2 v_r'} & \cdots & V_{v_r'}
    \end{array}
    \right)
\end{equation*}
We use linear uncertainty propagation to approximate ${\bf C}'$
\begin{equation}
    {\bf C}' \approx {\bf J}{\bf C}{\bf J}^{\rm T}
    \label{eq:cov_transformation}
\end{equation}
where ${\bf J}$ is the Jacobian matrix
\begin{equation*}
    {\bf J}\equiv\frac{\partial(\phi_1,\phi_2,\mu_{\phi1},\mu_{\phi2},\varpi',v_r')}{\partial(\alpha,\delta,\mu_\alpha,\mu_\delta,\varpi,v_r)}.
\end{equation*}

Note that $(\varpi',v_r')=(\varpi,v_r)$ in coordinate rotation. Also, other new coordinates do not explicitly depend on parallax and radial velocity. Therefore, we directly know $\partial\varpi'/\partial i = \delta_{i\varpi}$, $\partial v_r'/\partial i = \delta_{iv_r}$, $\partial i/\partial\varpi = \delta_{i\varpi'}$, and $\partial i/\partial v_r = \delta_{iv_r'}$, where the Kronecker symbol $\delta_{ij}=1$ only if $i=j$, otherwise 0. This greatly simplifies our calculation as we only need to take into account the rotation of angular coordinates ($\alpha$,$\delta$,$\mu_\alpha$,$\mu_\delta$) and ($\phi_1$,$\phi_2$,$\mu_{\phi1}$,$\mu_{\phi2}$) independent from $\varpi$ and $v_r$.

Rotation in the spherical coordinate system is nonlinear, which makes it challenging to derive ${\bf J}$ analytically. However, by first transforming the sky coordinates to the Cartesian system $(x,y,z,v_x,v_y,v_z)$, we can more easily deal with rotation as it becomes a linear coordinate transformation, which can be described by the $6\times6$ rotational matrix,
\begin{equation*}
    {\bf R}\equiv\left(
    \begin{array}{cc}
        {\bf R}_{3\times3} & {\bf 0} \\
        {\bf 0} & {\bf R}_{3\times3}
    \end{array}
    \right)
\end{equation*}
Here, ${\bf R}_{3\times3}\in {\rm SO}(3)$ is the standard 3D rotational matrix. After the rotation, we can transform the rotated Cartesian frame back to the great circle frame. The combined Jacobian matrix thus equals the product of the Jacobian matrices for the three transformations. Since rotation is a linear translation, the Jacobian matrix of rotation is simply ${\bf R}$ itself. We derive the two remaining Jacobian matrices as follows.

We consider the rotation of angular coordinates in the unit sphere. Such a simplification does not affect our calculation as the parallax and radial velocity are independent coordinates. The transformation from sky coordinates to the cartesian system is
\begin{equation*}
    \left(
    \begin{array}{c}
        x \\
        y \\ 
        z
    \end{array}
    \right) = \left(
    \begin{array}{c}
        \cos\alpha \cos\delta \\
        \sin\alpha \cos\delta \\ 
        \sin\delta
    \end{array}
    \right)
\end{equation*}
and
\begin{equation*}
    \left(
    \begin{array}{c}
        v_x \\
        v_y \\ 
        v_z
    \end{array}
    \right) = 
    \left(
    \begin{array}{c}
        -\mu_\alpha \sin\alpha - \mu_\delta \cos\alpha \sin\delta \\
        \mu_\alpha \cos\alpha - \mu_\delta \sin\alpha \sin\delta \\
        \mu_\delta \cos\delta
    \end{array}
    \right).
\end{equation*}
Recall that we define $\mu_\alpha\equiv\dot{\alpha}\cos\delta$. Therefore, the corresponding $6\times4$ Jacobian matrix is
\begin{equation*}
    {\bf J}_1\equiv\frac{\partial(x,y,z,v_x,v_y,v_z)}{\partial(\alpha,\delta,\mu_\alpha,\mu_\delta)}
    = \left(
    \begin{array}{cccc}
        -\sin\alpha \cos\delta & -\cos\alpha \sin\delta & 0 & 0 \\
        \cos\alpha \cos\delta & -\sin\alpha \sin\delta & 0 & 0 \\
        0 & \cos\delta & 0 & 0 \\
        -\mu_\alpha \cos\alpha+\mu_\delta \sin\alpha \sin\delta & -\mu_\delta \cos\alpha \cos\delta & -\sin\alpha & -\cos\alpha \sin\delta \\
        -\mu_\alpha \sin\alpha-\mu_\delta \cos\alpha \sin\delta & -\mu_\delta \sin\alpha \cos\delta & \cos\alpha & -\sin\alpha \sin\delta \\
        0 & -\mu_\delta \sin\delta & 0 & \cos\delta
    \end{array}
    \right).
\end{equation*}
We can also calculate the inverse transformation to the great circle frame,
\begin{equation*}
    \left(
    \begin{array}{c}
        \phi_1 \\
        \phi_2
    \end{array}
    \right) = \left(
    \begin{array}{c}
        \arctan\displaystyle\frac{y'}{x'} \\
        \arcsin z' \\ 
    \end{array}
    \right)
\end{equation*}
and
\begin{equation*}
    \left(
    \begin{array}{c}
        \mu_{\phi1} \\
        \mu_{\phi2}
    \end{array}
    \right) = \frac{1}{\sqrt{x'^2+y'^2}}\left(
    \begin{array}{c}
        -v_x' y' + v_y' x' \\
        v_z'
    \end{array}
    \right).
\end{equation*}
The $4\times6$ Jacobian matrix of this transformation is 
\begin{equation*}
    {\bf J}_2\equiv\frac{\partial(\phi_1,\phi_2,\mu_{\phi1},\mu_{\phi2})}{\partial(x',y',z',v_x',v_y',v_z')}
    = \left(
    \begin{array}{cccccc}
        -\displaystyle\frac{\sin{\phi_1}}{\cos{\phi_2}} & \displaystyle\frac{\cos{\phi_1}}{\cos{\phi_2}} & 0 & 0 & 0 & 0 \\
        0 & 0 & \displaystyle\frac{1}{\cos{\phi_2}} & 0 & 0 & 0 \\
        -\displaystyle\frac{\mu_{\phi 2} \sin{\phi_1} \sin{\phi_2}}{\cos{\phi_2}} & \displaystyle \displaystyle\frac{\mu_{\phi 2} \cos{\phi_1} \sin{\phi_2}}{\cos{\phi_2}} & 0 & -\sin{\phi_1} & \cos{\phi_1} & 0 \\
        0 & 0 & \displaystyle\frac{\mu_{\phi 2} \sin{\phi_2}}{\cos^2{\phi_2}} & 0 & 0 & \displaystyle\frac{1}{\cos{\phi_2}}
    \end{array}
    \right)
\end{equation*}
For clarity, we already write ${\bf J}_2$ in terms of the great circle coordinates. 

The combined Jacobian matrix is given by
\begin{equation*}
    {\bf J}=\left(
    \begin{array}{cc}
        {\bf J}_2{\bf R}{\bf J}_1 & {\bf 0} \\
        {\bf 0} & {\bf I}_{2\times2} \\
    \end{array}
    \right).
\end{equation*}
The identity matrix in the lower right accounts for the transformation of $(\varpi,v_r)$. We have verified that ${\bf J}_2{\bf R}{\bf J}_1$ also becomes the identity matrix in the case of no rotation, ${\bf R}={\bf I}_{6\times6}$. Finally, we can insert ${\bf J}$ to Eq.~(\ref{eq:cov_transformation}) to obtain the covariance matrix in the great circle frame.

\bibliography{references}
\bibliographystyle{aasjournalv7}

\end{document}